%% file: paper.tex
\documentclass[fleqn,usenatbib]{mnras}

\usepackage[T1]{fontenc}
\usepackage{ae,aecompl}
\usepackage{multirow}
\usepackage{dirtytalk}

\usepackage[normalem]{ulem}

\usepackage{graphicx}	
\usepackage{amsmath}	
\usepackage{amssymb}	
\usepackage{pdfpages}   
\usepackage{hyperref}

\graphicspath{{figures/}}

\newcommand{\dd}{\text{d}}

\newcommand{\avg}[1]{\langle #1 \rangle}

\newcommand{\new}[1]{{#1}}
\newcommand{\newer}[1]{{#1}}

\newcommand{\PreFRBLE}{{\sc PrEFRBLE}}

\newcommand{\frbpoppy}{{\sc frbpoppy}}

\newcommand{\unitDM}{{\rm\ pc \ cm^{-3}}}

\newcommand{\DMobs}{\text{DM}_{\rm obs}}
\newcommand{\DMEG}{\text{DM}_{\rm EG}}
\newcommand{\DMMW}{\text{DM}_{\rm MW}}
\newcommand{\DMIGM}{\text{DM}_{\rm IGM}}

\newcommand{\DM}{\text{DM}}

\newcommand{\fIGM}{f_{\rm IGM}}
\newcommand{\fIGMzero}{f_{\rm IGM,0}}
\newcommand{\zFRB}{z_{\rm FRB}}

\newcommand{\Bayes}{\mathcal{B}}

\newcommand{\arxiv}[1]{ #1 } 

\title[PrEFRBLE]{Joint inference on the redshift distribution of fast radio burst and on the intergalactic baryon content}

\author[Hackstein et al.]{
{S. Hackstein$^{1}$}\thanks{E-mail: stefan.hackstein@hs.uni-hamburg.de},
{M. Br\"uggen$^{1}$},
{F. Vazza$^{1,2}$},
\\
$^{1}$Hamburger Sternwarte, University of Hamburg, Gojenbergsweg 112, 21029, Germany\\
$^{2}$University of Bologna,  Department of Physics and  Astronomy,
Via Gobetti 93/2, I-40129, Bologna, Italy;\\$^{\phantom{22}}$Istituto di Radioastronomia, INAF, Via Gobetti 101,40129 Bologna, Italy\\
}
\date{Accepted 2020 December 18. Received 2020 December 17; in original form 2020 August 25}

\begin{document}
\label{firstpage}
\pagerange{\pageref{firstpage}--\pageref{lastpage}}
\maketitle

\begin{abstract}

\arxiv{\textit{Context}:} Fast radio bursts are transient radio pulses of extragalactic origin.
Their dispersion measure is indicative of the baryon content in the ionized intergalactic medium between the source and the observer.  
However, inference using unlocalized fast radio bursts is degenerate to the distribution of redshifts of host galaxies. 
\arxiv{

\textit{Method:}} We perform a joint inference of the intergalactic baryon content and the fast radio burst redshift distribution with the use of Bayesian statistics by comparing the likelihood of different models to reproduce the observed statistics in order to infer the most likely models.
\new{In addition to two models of the intergalactic medium, we consider contributions from the local environment of the source, assumed to be a magnetar, as well as a representative ensemble of host and intervening galaxies.}
\arxiv{

\textit{Results}:} 
Assuming that the missing baryons reside in the ionized intergalactic medium, our results suggest that the redshift distribution of observed fast radio bursts peaks at $z \lesssim 0.6$.
\newer{However, conclusions from different instruments regarding the intergalactic baryon content diverge
and thus require additional changes to the observed distribution of host redshifts, beyond those caused by telescope selection effects.}
\end{abstract}





\input{introduction}

\input{method}

\input{results}

\input{discussion}

\input{conclusions}


\section*{Acknowledgements}
This work was funded by the Deutsche Forschungsgemeinschaft (DFG) under grant BR2026/25.

SH thanks Rainer Beck, Laura Spitler and Charles Walker for interesting and fruitful discussions on many aspects of this work.

Our cosmological simulations were performed with the {\sc ENZO} code (http://enzo-project.org), under project HHH38 and HHH42 at the J\"ulich Supercomputing Centre (P.I. F. Vazza). 
FV acknowledges financial support from the ERC  Starting Grant \say{MAGCOW}, no. 714196.

We further acknowledge the use of computational resources at the Rechenzentrum of the University of Hamburg.

\section*{Data availability}

The data underlying this article are available in the PrEFRBLE repository, at \href{http://doi.org/10.5281/zenodo.3862636}{http://doi.org/10.5281/zenodo.3862636}.




\bibliographystyle{mnras}
\bibliography{cites} 





\label{lastpage}
\end{document}

%% file: introduction.tex
\section{Introduction}

Observations of the cosmic microwave background (CMB) show that $\approx 5 \%$ of energy density in the Universe exists as ordinary matter \citep{Planck2014}.
However, in the  $z \leq 2$ range,  observations of stars and gas in galaxies, of the hot intracluster medium and of the Ly$\alpha$-forest can only account for about half of that amount \citep[e. g.][]{nicastro2008}.
The "missing" baryons are believed to be hidden in the warm-hot intergalactic medium (WHIM) \citep{cen1999cosmic,dave2001baryons}, with temperatures of $10^5 - 10^7 ~\rm K$ and low baryon densities of $10^{-6} - 10^{-5} ~\rm cm^{-3}$.
This medium is difficult to observe directly as it only couples with radiation through electronic transitions.
However, correlation of the distribution of galaxies with CMB distortions due to Sunyaev-Zel'dovich effect has revealed the likely presence of filaments of warm-hot gas between galaxies, possibly consistent with the missing baryons \citep[e. g.][]{tanimura2018,deGraaff2019}.
\\

Several studies have suggested that fast radio bursts (FRBs) can be used to detect the missing baryons.
Their signals propagate across cosmological distances and get dispersed by diffuse ionized gas, quantified by the dispersion measure (DM), defined as the column density of free electrons
\begin{equation}
    \DM = \int n_e \dd l .
\end{equation}
\citet{wei2019} proposed to compare the redshift evolution of the DMs using $\approx$ 3000 FRBs with known redshifts to that of the Hubble parameter to measure the IGM baryon content $\fIGM$, i. e. the amount of all baryons that reside in the ionized IGM.
\cite{deng2014} argued that the redshift observed for $\gamma$-ray bursts associated to FRBs can be used to infer $\fIGM$. 
\cite{mcquinn2013} modelled the distribution of DM expected for FRBs from different redshift and found that $\sim 100$ DM of FRBs from redshift $z>0.5$, localized with sub-arcminute precision around identified galaxies, can be used to infer the baryon profile of galaxies in order to constrain the localization of cosmic baryons.
\cite{munoz2018} proposed to cross-correlate the DM of arcminute-localized FRBs with the thermal Sunyaev-Zeldovich effect in the CMB.
The latter depends on the temperature of the WHIM  and the amount of baryons localized in the WHIM, which can be constrained using $\sim 1000$ FRBs with arcminute localization.
\cite{qiang2020reconstructing} suggest to use model-independent Gaussian processes to investigate the evolution of $\fIGM$ with DM of FRBs with identified redshift.
\\

All of the aforementioned papers require a large number of well-localized FRBs or cross-correlation with associated observations.
However, \cite{yang2020} argue that the majority of FRBs are not supposed to have strong associated persistent sources.
Furthermore, the exact localization of sources of short-duration signals without known redshift is not trivial to constrain, and neither is the identification of the host galaxies of FRBs \citep{eftekhari2017,mahony2018,marcote2019,prochaska2019uncovering}.
The host can be identified using interferometry or by observing persistent counter parts to the FRB.
So far, FRBs have been located to a multitude of galaxy types, from star-forming dwarf galaxies \citep{tendulkar2017host} to very massive galaxies with old stellar population \citep{ravi2019fast,bannister2019single}.
Unluckily, the small sample of localized FRBs cannot provide enough information to arrive at reasonable conclusions regarding cosmological questions.
However, in this work, we show how it is possible to use DM of a large sample of unlocalized FRBs,  in order to infer their host redshift distribution in a statistical way, which also allows us to constrain the amount of ionized baryons located in the cosmic web.  
\\

Because of the unknown distance to unlocalized FRBs, the assumed distribution of source redshifts can help interpret the distribution of observed DM.
Several papers have tried to infer the intrinsic redshift distribution of FRBs,  either by modelling the distribution of DM and other FRB properties with analytical or Monte-Carlo methods \citep{bera2016,caleb2016,gardenier2019synthesizing}, or by performing a luminosity-volume test  \citep{locatelli2018}.
So far there are inconsistencies in implications of data from different instruments, e. g. ASKAP requiring faster change in FRB density with redshift than Parkes.

\citet{macquart2018fluence} use the flux density and fluence of FRBs to infer the luminosity distance and evolutionary history, as well as the redshift distribution, and show how the history of the ionized IGM affects the distribution of observed DM, thus influences our inference of the host redshift distribution.
\\

Here we propose to use unlocalized FRBs to perform a joint analysis of the FRB redshift distribution and the IGM baryon content,  $\fIGM$, by comparing the expected distribution of DM to the available observed values reported by Parkes, CHIME and ASKAP observatories \citep{Parkes,macquart2010,CHIME/FRB2018}, using the \PreFRBLE\footnote{\href{https://github.com/FRBs/PreFRBLE}{github.com/FRBs/PreFRBLE}} software \citep{PreFRBLEzenodo}, presented in \citet{hackstein2020prefrble}.
In Sec. \ref{sec:method} we explain how we model different values of $\fIGM$.
The resulting predictions are presented and compared in Sec. \ref{sec:results}.
We discuss our results in Sec. \ref{sec:discussion} and conclude in Sec. \ref{sec:conclusion}.

%% file: method.tex
\section{Method}
\label{sec:method}


Investigation of the fraction of baryons in the ionized IGM, $\fIGM = \Omega_{\rm IGM} / \Omega_{\rm baryons}$, with DM requires comparison of the observed DM with expectations of $\DM(z|\fIGM)$, according to source redshift $z$, \citep{keane2016host}
\begin{equation}
\avg{\DM(z)} = \frac{c \rho_{\rm crit} \Omega_b }{ m_p\mu_e H_0 }\int \fIGM(z) \frac{(1+z)}{H(z)} ~\dd z .
\label{eq:DM_cosmo}
\end{equation}
However, FRBs do not come with a direct measure of redshift and for the majority of FRBs, that could not be localized by other means, the DM is the best indicator for the source distance \citep{dolag2015,Niino2018,luo2018,Walker2018,Pol2019}.
Still, by assuming different plausible redshift distributions of FRBs, $\pi(z)$, we can estimate the distribution of DM to be observed by instruments, which is determined by $\fIGM$.

By comparing the expected distribution to observations, we can quantify the likelihood of different combinations of $\pi(z)$ and $\fIGM$.
Consequently, this allows us to put constraints on the WHIM density without the need to localize FRBs.
Furthermore, by a detailed investigation of the redshift-evolution of $\fIGM$ in cosmological simulations, the inference presented here can be used to investigate the helium-reionization history \citep[e. g.][]{linder2020,dai2020reconstruction}.
\\

In \citet{hackstein2020prefrble}, we obtained likelihood estimates, $L(\DM|z)$, to observe extragalactic DM from source at redshift $z$. 
These expectations consider contributions from all regions along the LoS in our benchmark scenario, considering magnetars as the source of FRBs \citep{Piro2018} as well as a realistic ensemble of different host and intervening galaxies \citep{lacey2016unified,rodrigues2018}.
Predictions for the IGM were obtained using a constrained cosmological simulation of the local Universe, produced using the cosmological magneto-hydrodynamic code ENZO \citep{Bryan2014} together with initial conditions obtained following \citet{Sorce2016} and cosmological parameters of PLANCK \citep{Planck2015params}.
The constrained volume of $250 ~\rm (Mpc/h)^3$ was embedded in a $500 ~\rm (Mpc/h)^3$ grid of $1024^3$ cells.
The use of five adaptive mesh refinement levels allowed for a maximum resolution of $30 ~\rm kpc$ in the most-dense environments.
The simulation starts at redshift $z=60$, where all baryons are in the IGM.
However, the limited resolution does not allow us to properly resolve galaxy formation and the condensation of cold gas out of the IGM, thus implies $\fIGM=1$, always.
Further information on this model can be found in \citet{hackstein2018} and \citet{hackstein2019}.
A reduced version of this model can be found on \href{https://crpropa.desy.de}{crpropa.desy.de} under \say{additional resources}, together with the other models presented in \citet{hackstein2018}.

\newer{The contribution of the circumburst environment of magnetars is modelled using the analytical predictions of the wind model in \citet{Piro2018} in a Monte-Carlo simulation.
More details on this simulation and the assumed priors can be found in \citet{hackstein2019}.
The contribution of host and intervening galaxies is estimated by computing LoS integrals in a realistic ensemble of galaxy models that recreate a wide variety of properties of the population of galaxies in the Universe \citep{lacey2016unified,rodrigues2018}. 
More details on this Monte-Carlo simulation and the resulting likelihoods can be found in \citet{hackstein2020prefrble}.}

We further use \frbpoppy\footnote{\href{https://github.com/davidgardenier/frbpoppy}{github.com/davidgardenier/frbpoppy}} \citep{gardenier2019synthesizing} to model the distribution of host redshift of observed FRBs, $\pi(z)$.
We assume three physically motivated intrinsic redshift distributions, i. e. following the comoving volume (coV), stellar mass density (SMD) or star formation rate (SFR), together with the selection effects of ASKAP, CHIME and Parkes instruments.
\new{We use the parameters of their \textit{complex} model, which best reproduces observations.}

Together, these can be used to quantify the distribution of expected DMs, 
\begin{equation}
L(\DM) = \int L(\DM,z)\times \pi(z) ~\dd z.
\label{eq:tele-DM}
\end{equation}
For more details on the derivation of likelihood $L(\DM|z)$ and prior $\pi(z)$, see \citet{hackstein2020prefrble}.
\\

Numerical simulations suggest that up to $28 \pm 11$ per cent of baryons are located in neutral hydrogen clouds in the IGM, observable in the photoionized Ly$\alpha$ forest, as well as $25 \pm 8$ per cent in the WHIM, leaving a baryon shortfall of $29 \pm 12$ per cent \citep{shull2012}.
This leaves a plausible range of $0.3 \lesssim \fIGM \lesssim 0.9$ \citep[cf.][]{li2019}.

As explained above, in the constrained simulation $\fIGM$ is assumed to be constant with redshift, thus to be a global factor to $\avg{\DM(z)}$ (Eq. \ref{eq:DM_cosmo}).
This allows to obtain estimates for different values of $\fIGM$ in post-processing, allowing us to test how many  missing baryons can be located in the WHIM \citep{deng2014,keane2016host}.
To this end, we modify the likelihood of intergalactic contribution, $L(\DMIGM|z)$, obtained from the constrained simulation of the IGM, such that the result returns the required average value according to Eq. \ref{eq:DM_cosmo}, while shape and norm are conserved:
\begin{equation}
L(\DMIGM|\fIGM) = \fIGM \times L(\fIGM\times\DMIGM) .
\label{eq:shift_fIGM}
\end{equation}
\\

\new{
The fraction of baryons in the IGM,  $\fIGM$, is believed to grow with redshift \citep{mcquinn2013,Prochaska2019}.
 Simulations suggest that $\fIGM \approx 0.9$ prior to $z=1.5$ \citep{meiksin2009physics}.
To account for the redshift evolution of $\fIGM$, we consider a model for the IGM following \citet{pshirkov2016} (called Pshirkov16 hereafter).
We perform a Monte-Carlo simulation in which we divide the LoS in segments of Jeans-length size 
$\lambda_J(z) = 2.3 ~ (1+z)^{-1.5} ~\rm Mpc$.
Assuming that the electron density $n_e$ is well described by the distribution of neutral hydrogen deduced from the observed Lyman-$\alpha$ forest, we pick random values for comoving $n_e$ drawn from a log-normal distribution of overdensity according to Eq. (2) in \citet{pshirkov2016}.
Finally, $n_e$ is multiplied by the value of $\fIGM$ at corresponding redshift.
We assume $\fIGM$ to increase linearly with redshift from $\fIGM=\fIGMzero$ at $z=0$ to $\fIGM = 0.9$ at $z=1.5$, while it stays at this constant value at higher redshifts.
}
\\

In order to quantify the likelihood of a scenario of $\fIGM$ and FRB redshift distribution $\pi(z)$ to account for the observation of an individual FRB with $\DMobs = \DMEG + \DMMW$, we take the value of $L(\DMEG|\fIGM)$ for the benchmark scenario, using the estimates for $\DMMW$ according to NE2001 model \citep{NE2001} as listed in FRBcat \citep{FRBCAT}.
These estimated values of $\DMEG$ were shown to be correct to $\approxeq 30 \unitDM$ \citep{manchester2005australia}.

In order to compare different choices of $\fIGM$ and host redshift distribution, we compute the Bayes factor 
\begin{equation}
    \Bayes( \DMEG | f_{{\rm IGM},i}, f_{{\rm IGM,ref}} ) =\frac{L(\DMEG|f_{{\rm IGM},i})}{L(\DMEG|f_{{\rm IGM,ref}})} 
\label{eq:Bayes}
\end{equation}
by renormalizing all likelihoods of individual FRBs to the same reference choice of $f_{{\rm IGM,ref}}$.
$\Bayes$ quantifies how much the observation of $\DMEG$ changes our corroboration in favour of $f_{{\rm IGM},i}$ compared to $f_{{\rm IGM,ref}}$.
By renormalizing all $\Bayes$ to the best-fit scenario $\Bayes_{\rm max}$, the presented values quantify how much the observations change our corroboration in disfavour of the corresponding scenario.

For FRBs with localized hosts, the information on $\fIGM$ and $\pi(z)$ can be distinguished, thus allowing for separate likelihoods.
For the events listed with host redshift $\zFRB$ in FRBcat, instead of the procedure explained above, we obtain the likelihood of $\fIGM$ by comparing values of $L(\DMEG|\zFRB,\fIGM)$ and compute the Bayes factor $\Bayes$ by renormalizing the likelihood to a particular choice of $f_{\rm IGM,i}$. Likewise, we compare values of $\pi(\zFRB)$ for their implication on the redshift distribution.
\\

$\Bayes$ obtained for individual events and different instruments can simply be multiplied in order to interpret the combined results. 
Note that $\Bayes$ quantifies how much the current observations change our corroboration between the considered scenarios.
However, according to Bayes theorem
\begin{equation}
P(\fIGM|\DMEG) \propto L(\DMEG|\fIGM) \pi(\fIGM) ,
\label{eq:Bayes-theorem}
\end{equation}
the ratio of likelihoods $L$, $\Bayes$, has to be multiplied by the ratio of priors $\pi$ in order to arrive at the ratio of posteriors $P$.
$\pi(\fIGM)$ quantifies how likely different values of $\fIGM$ are prior to investigation of $\DMEG$, while $P$ expresses our conclusion on the likelihood of values of $\fIGM$ after consideration of the data.
However, we consider no preference for any value of $\fIGM$, $\pi(\fIGM) = \text{const.}$, thus $\Bayes$ is identical to the ratio of $P$.

%% file: results.tex

\section{Results}
\label{sec:results}
\begin{figure*}
    \centering
    {\large constrained simulation, $\fIGM=const.$}
    
    \includegraphics[width=\textwidth]{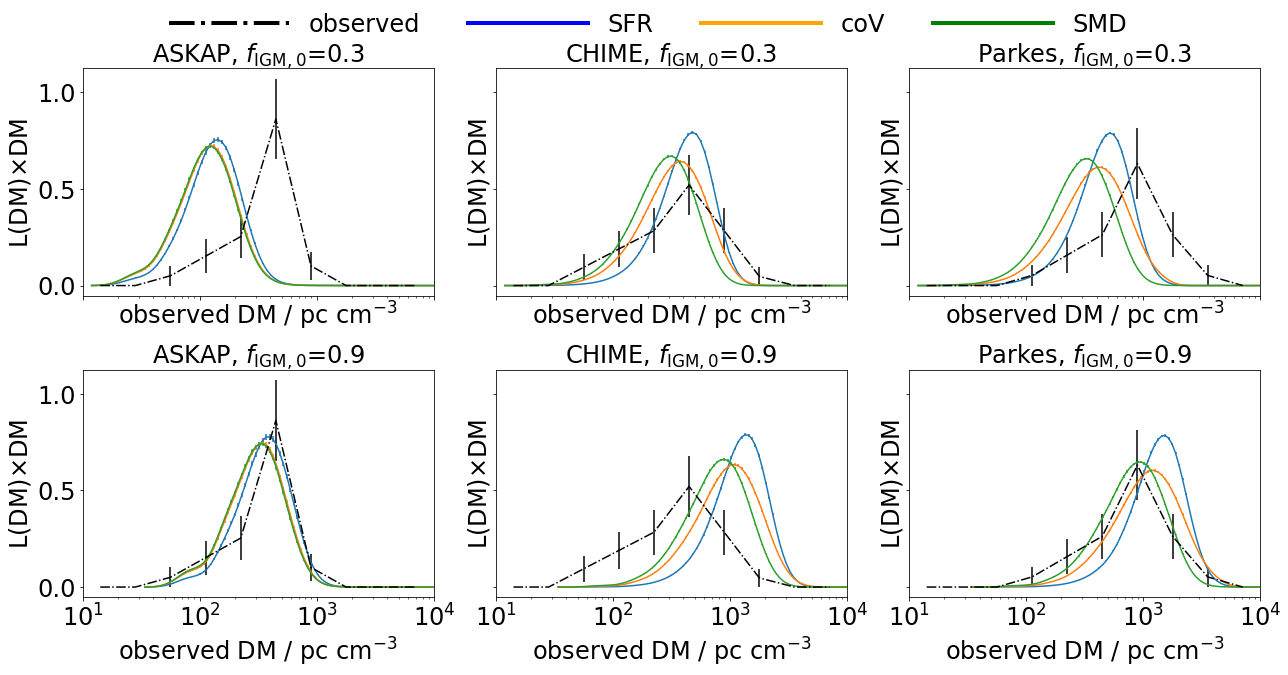}

    \caption{Distribution of extragalactic $\DMEG = \DMobs - \DMMW$ as observed (dash-dot) by ASKAP (left), CHIME (center) and Parkes (right) minus estimate of MW contribution based on NE2001 model, as listed in FRBcat.
    The solid lines show the expected distribution to be observed by the corresponding instrument according to Eq. \ref{eq:tele-DM}, assuming FRBs from magnetars in our benchmark scenario using the constrained simulation to model the IGM and extreme constant values for the baryon content, $\fIGM=0.3$ (top) or $\fIGM=0.9$ (bottom), as well as FRB redshift distribution to follow star formation rate (SFR, blue), comoving volume (coV, orange) or stellar mass density (SMD, green).
    The error bars show the shot noise of the observed data.
    For the expected distribution, the barely visible error bars are the shot noise of the Monte-Carlo samples used to obtain the likelihood function.
    The product of Bayes factors (Eq. \ref{eq:Bayes}) for individual FRBs, shown in Fig. \ref{fig:joint_telescopes}, quantifies how well different scenarios recreate the observed distribution.
    }
    \label{fig:DM_telescopes}

    {\large Pshirkov+16, $\fIGM\propto z$}
    
    \includegraphics[width=\textwidth]{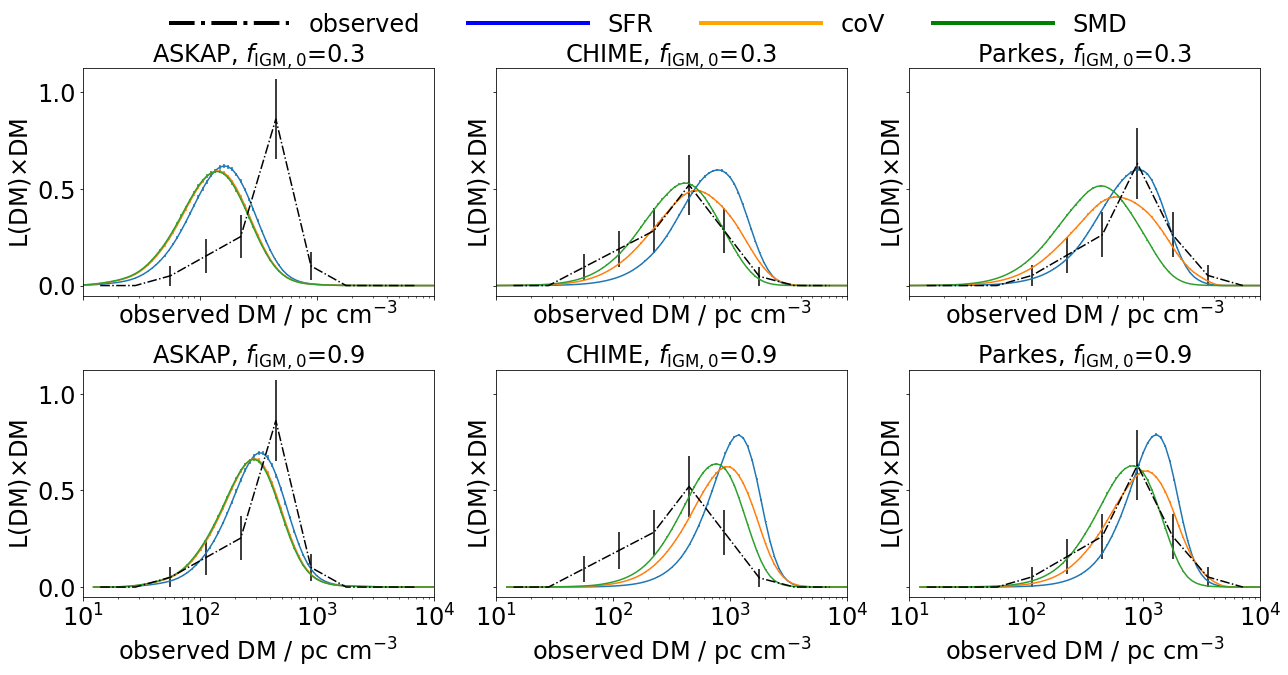} 
    \caption{ Same as Fig. \ref{fig:DM_telescopes}, using the Pshirkov16 model with redshift dependent $\fIGM$ for the IGM and extreme values for its present value $\fIGMzero$.
    Models for the redshift distribution, circumburst environment, host and intervening galaxies remain the same.}
    \label{fig:DM_telescopes_pshirkov}
\end{figure*}

\subsection{Expected distribution vs. observation}

In Fig. \ref{fig:DM_telescopes} we show the results for all combinations of instruments and redshift distributions for two extreme constant values of $\fIGM=0.3$ and $\fIGM=0.9$, using the constrained simulation for the IGM.
These can be compared directly to observations by the corresponding instrument, which are indicated by the dash-dotted line.

The majority of FRBs observed by ASKAP have $\DMEG\approx 500 \unitDM$.
The expected peak values of $\DMEG$ for $\fIGM$ between 0.3 and 0.9 range from $100 \unitDM$ to $440 \unitDM$.
ASKAP results are thus in favour of high values of $\fIGM\approx 0.9$.
However, for the values of DM observed with ASKAP that are generally higher than expectations, other foregrounds such as the local contributions need to be considered more carefully in order to arrive at reasonable conclusions.
The assumed intrinsic redshift distributions do not differ much in their predictions at low $z$ \citep[cf.][]{hackstein2020prefrble}, thus rendering ASKAP results incapable of distinguishing FRB populations with a small sample of FRBs.

While the distribution of DM observed by CHIME peaks at roughly the same value as for ASKAP $\DMEG \approx 500 \unitDM$, the peak is less pronounced and the upper $1\sigma$-deviation is more than 1.5 times higher than for ASKAP, reaching values beyond $1000 \unitDM$.
However, for CHIME, the expected peak values of $\DMEG$ range from $300 \unitDM$ to $1500 \unitDM$ for $\fIGM=0.3$ and $\fIGM=0.9$, respectively, far above the observed value, hence expecting many more high values of DM.
Thus, the CHIME results favour low $\fIGM \gtrsim 0.3$, depending on the assumed distribution of the host redshifts.
Also, CHIME observes more FRBs with $50 < DM_{\rm EG} ~/ \unitDM < 200$ than ASKAP, which is not expected according to results in Fig. \ref{fig:DM_telescopes}.
This hints on an imprecise assumption on the redshift distribution of FRB sources visible to CHIME, that observes at different wavelengths than the other two instruments.

For Parkes, the expected peak value of $\DMEG$ ranges from $300 \unitDM$ to $1800 \unitDM$ between $\fIGM=0.3$ and $\fIGM=0.9$, similar as expected for CHIME.
However, the observed sample peaks at $\DMEG \approx 1000 \unitDM$ - about twice the value observed with CHIME - which is well within the expected range.
Hence, different assumed redshift distributions will favour other values of $\fIGM$, e. g. a population concentrated at low redshifts, such as SMD, will need a high value of $\fIGM\lesssim 0.9$ to explain the observed DMs, while populations with greater redshifts, such as SFR, will require lower values of $\fIGM$.
The Parkes sample is dominated by high values of DMs, with $\approx 30$ per cent of FRBs observed with $\DM > 1000 \unitDM$.
The majority of these is thus likely located at $z \gtrsim 1$, such that local contributions to the DMs are less significant. This makes the Parkes results overall the most reliable for investigation of $\fIGM$ and other cosmological properties.
\\

\new{
In Fig. \ref{fig:DM_telescopes_pshirkov} we present the results for all combinations of instruments and redshift distributions for two extreme values of $\fIGMzero=0.3$ and $\fIGMzero=0.9$ at $z=0$, by modelling the IGM with the Pshirkov16 model, including the linearly increasing $\fIGM$.
For the case that $f_{{\rm IGM},0} = 0.9$, $\fIGM = \rm const.$ at all redshifts, similar to the constrained simulation in the lower row of Fig. \ref{fig:DM_telescopes}.
The fact that predictions are hardly distinguishable verifies the similarity of both models \footnote{A comparison of average $\langle \DM \rangle(z)$ can be found in \citet{hackstein2020prefrble}}
However, for lower values of $\fIGMzero$ there is stronger contribution for $\DM$ at large redshift, thus larger values of $\DM$ are predicted in this case.
This reduces the variance of predictions with different $\fIGMzero$, when compared to varying constant $\fIGM$.
}


\subsection{Inference of $\fIGM$ and $\pi(z)$}
\begin{figure*}
    \centering
    {\large constrained simulation, $\fIGM=const.$}

    \includegraphics[width=0.3\textwidth]{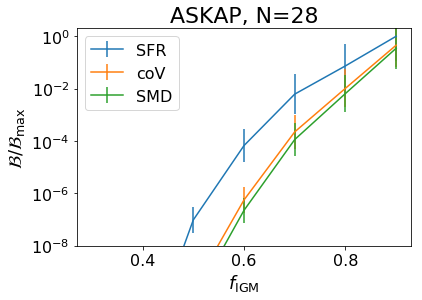}
    \includegraphics[width=0.3\textwidth]{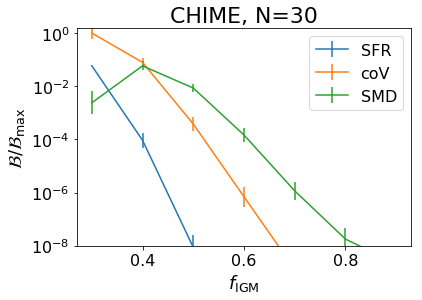}
    \includegraphics[width=0.3\textwidth]{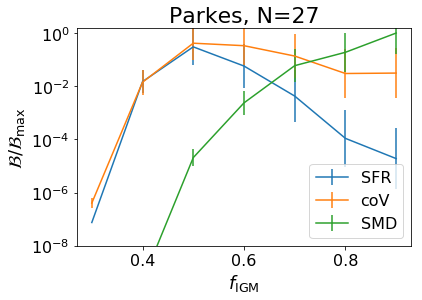}

    \caption{Bayes factor $\Bayes$ of FRBs found in FRBcat comparing scenarios for different baryon content in IGM $\fIGM$ and redshift distributions of FRBs, following either star formation rate (SFR), comoving volume (coV) or stellar mass density (SMD), assuming that FRBs are produced by magnetars in our benchmark scenario.
    To model the IGM we use the constrained simulation described in detail in \citet{hackstein2020prefrble}.
    The title indicates the instrument that observed the considered number of $N$ FRBs.
    The estimated value and deviation are obtained via Jackknife resampling.
    All $\Bayes$ are renormalized to the best-fit $\Bayes_{\rm max}$ to indicate how much less likely a given scenario reproduces the observed data (cf. Fig. \ref{fig:DM_telescopes}).
    }
    \label{fig:joint_telescopes}
    {\large Pshirkov16, $\fIGM\propto z$}

    \includegraphics[width=0.3\textwidth]{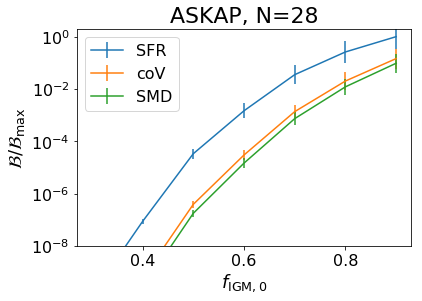}
    \includegraphics[width=0.3\textwidth]{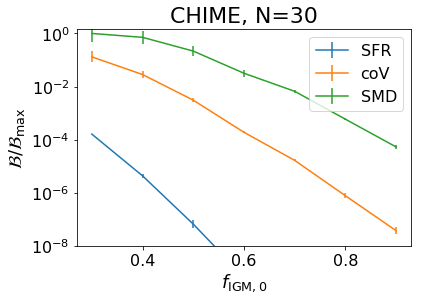}
    \includegraphics[width=0.3\textwidth]{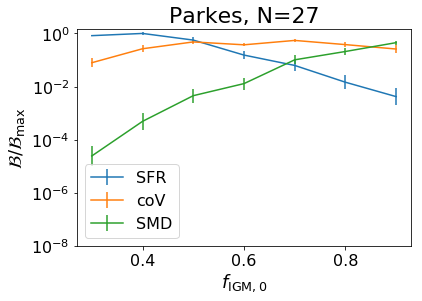}
    \caption{Same as Fig. \ref{fig:joint_telescopes}, using the Pshirkov16 model for the IGM contribution.
    Models for the redshift distribution, circumburst environment, host and intervening galaxies remain the same.}
    \label{fig:joint_telescopes_Pshirkov}
\end{figure*}

\subsubsection{$\fIGM$ from localized FRBs}
Recently, \citet{li2020cosmologyinsensitive} estimated $\fIGM=0.84^{+0.16}_{-0.22}$ using DM of the five localized FRBs together with the corresponding luminosity distance of Ia supernovae.
Shortly after that, \citet{bhandari2020host} report the localization of two additional FRBs.
Using their reported values of $\DMEG = \DMobs - \DMMW$ for all six FRBs localized at decent redshift, $z\geq 0.1$, we estimate $\fIGM = 0.84^{+0.11}_{-0.39}$ within 3$\sigma$ using our more sophisticated approach \new{with our benchmark scenario and the Pshirkov16 model for the IGM.
The lower limit suggests that $\fIGM \gtrsim 0.5$.
The current sample of localized FRBs is thus not yet sufficient to solve the missing baryons problem unambiguously \citep[cf.][]{Macquart_2020}.}

Though our inference depends on the chosen cosmology \citep{Planck2015params}, as the models for IGM as well as host and intervening galaxies in our benchmark scenario are obtained from a specific cosmological simulation, we derive the same estimated value, with a more conservative lower limit.
Still, the conclusion that this confirms the location of missing baryons in the WHIM should be taken with caution, at least because $\fIGM$ is considered to be constant with redshift.
Note that the likelihood function used in this estimate for $\fIGM$, i. e. $\prod\limits_i L(\fIGM|\DM_i,z_i)$, can be used as prior to infer the host redshift distribution using unlocalized FRBs, replacing assumptions on $\fIGM$.

\subsubsection{$\fIGM$ from individual instruments}

In Fig. \ref{fig:joint_telescopes} we show the Bayes factors $\Bayes$ resulting from the joint analysis of the FRB redshift distribution and of the IGM baryon content $\fIGM$, \new{assumed to be constant in our constrained simulation}.
For the 28 FRBs observed by ASKAP, high values of $0.8 \leq \fIGM$ are strongly favoured, regardless of the assumed redshift distribution.
Hence, ASKAP values seem to confirm the location of missing baryons in the ionized IGM, most probably the WHIM.
Due to the generally low distance of ASKAP events, $\Bayes$ shows the lowest variability between the assumed host redshift distributions.
However, estimates using ASKAP results are more vulnerable to the unknown contribution of the host galaxy and local environment of the source and should be taken with great caution.
This is further stressed by the fact that ASKAP likely prefers too high $\fIGM \gtrsim 0.9$. 

The 30 FRBs observed by CHIME, of which 18 have been found to repeat \citep{amiri2019,andersen2019,fonseca2020}, clearly favour low values of $\fIGM<0.5$ as  the observed distribution of DM concentrates on lower values than expected.
Since SFR redshift distribution peaks at a higher redshift, $z \approx 1$, it is disfavoured by CHIME results, except for the case of $\fIGM \approx 0.3$.
Instead, the SMD distribution, which peaks at lowest redshift, $z \approx 0.5$ is preferred for all probed values of $\fIGM > 0.4$ and favours $\fIGM = 0.4$.
Only for $\fIGM \lesssim 0.4$, coV and SFR show higher values of $\Bayes$. 
However, the parameters for CHIME selection effects available in \frbpoppy~ are an early version \citep[see][]{gardenier2019synthesizing}, hence results for this instrument should be considered with caution.
Still, if the current parameters are proven to be at least roughly correct, the present results either indicate a wrong assumption entering the expected distribution of FRB redshifts, which is likely as CHIME observes a different frequency band than ASKAP and Parkes, or suggest a very low baryon content in the IGM,  which we deem as unplausible, as it would contradict the result of any cosmological simulation.
However, CHIME results so far seem to disfavour the "WHIM solution" to the missing baryons problem. 

The 27 FRBs observed by Parkes show a high variation among the assumed host redshift distributions.
The SFR redshift distribution peaks at highest redshift $z \approx 1.1$ and favours $0.4\lesssim\fIGM<0.7$, \new{while it cannot be brought into agreement with the WHIM solution, as $\Bayes( \fIGM > 0.7 ) < 10^{-2}$ indicates that these values can be ruled out with more than 99 per cent certainty}.
The SMD distribution, peaking at $z \approx 0.6$ prefers a higher $\fIGM \geq 0.7$ value. 
The coV distribution peaks at $z\approx 0.7$ and favours $\fIGM \approx 0.6$, however, with relativley low significance.

\new{
In Fig. \ref{fig:joint_telescopes_Pshirkov} we show the Bayes factors $\Bayes$ resulting from the joint analysis of the FRB redshift distribution and of the IGM baryon content $\fIGMzero$ at $z=0$, assumed to increase linearly until $\fIGM=0.9$ at $z=1.5$ in the Pshirkov16 model.
Since predictions show less variance in $\fIGMzero$, the graphs are shallower than in Fig. \ref{fig:joint_telescopes}, while the general trends remain the same.
ASKAP results are in favour of the WHIM solution to the missing baryon problem.
However, CHIME results strongly disfavour high densities of baryons in the local IGM, for all assumed redshift distributions.
In particular, there is no value of $\fIGMzero$ for which $\Bayes > 10^{-2}$ with both, ASKAP and CHIME.
Such results can thus not be interpreted consistently, without considering further features that differ between the telescopes, e. g. systematic differences in the local contribution or the population of potentially visible FRBs.
However, investigation of possible solutions is beyond the scope of this paper.
}
\\

Not only does the inference of the different data sets differ dramatically, but also the assumed redshift distribution of FRBs strongly affects our interpretation of the data.
Thus, the degenerate problem of inferring $\fIGM$ with unlocalized FRBs is unlikely to be solved simply by increasing the sample size.
This stresses the importance of careful investigation of all assumptions entering the interpretation of measures of FRBs in order to arrive at reasonable conclusions.
As for now, we cannot derive strong or unambiguous conclusions on $\fIGM$ and on the intrinsic redshift distribution from unlocalized FRBs.
Only for the CHIME results we can conclude that the underlying redshift distribution has to peak at $z\lesssim 0.6$, in order to explain the high number of sources with low DMs. 
This conclusion is independent of the details of the assumed intrinsic redshift distributions and selection effects.

%% file: discussion.tex

\section{Discussion}
\label{sec:discussion}

Our work highlights the fact that the distributions of FRBs 
obtained by different radio telescopes lead to very different inferences about the cosmological distribution of such events.
The disagreement between ASKAP and CHIME is to be explained by a strong local contribution of DM to the bright low-distance FRBs observed by ASKAP.
These would account for systematically increased values of the observed DM.
By not accounting for such a strong local contribution in probability estimates, the increase in DM would be misinterpreted as IGM contribution, calling for a high $\fIGM$.
If the high DMs observed with ASKAP can be attributed to a stronger local contribution, lower values of $\fIGM$ would be favoured by FRB observations. This would reduce the fraction of missing baryons found in the WHIM.
\\

However, the fact that CHIME sees a higher amount of FRBs with $50 < \DM ~/ \unitDM < 200$ than ASKAP cannot be explained by low baryon content in the IGM or the general FRB population to be concentrated at low redshifts, which would affect CHIME and ASKAP in the same way.
If this feature persists in larger samples, it provides a strong hint towards CHIME observing a different subset of FRB sources than ASKAP.
The more than 10 times higher gain of CHIME allows it to potentially measure a different set of fainter sources \citep{amiri2018}, e. g. magnetars of older age \citep[cf.][]{metzger2017,marcote2020}.
Since these fainter sources are only visible at low redshift, this would increase the likelihood for CHIME to observe FRBs with low DM.
Furthermore, the local environment of older magnetars contributes much less to observed DM \citep{Piro2018}.
Thus, old magnetars could account for the increased number of FRBs observed with low DM compared to our expectations that consider the same population of magnetars to be observed by CHIME and ASKAP.
Such an increase in the likelihood to observe FRBs from low redshift would also make the CHIME results favour higher values of $\fIGM$.
However, since CHIME is also able to observe FRBs visible with Parkes \citep[e. g. FRB121102, see][]{josephy2019}, we require CHIME to detect more events with $\DM > 800 \unitDM$ and any above $> 2000 \unitDM$ in the future. 
If there is an extended set of sources available, this should reflect in a higher rate of FRBs observed by CHIME than extrapolated from other surveys \citep[e. g.][]{chawla2017}.

Note that 18 of the 30 FRBs observed by CHIME we considered here are repeaters \citep{amiri2019,andersen2019,fonseca2020}.
The high number of repeaters already observed by CHIME would suggest an increased chance to observe fainter bursts from repeating sources, e. g. from old magnetars.
For these young neutron stars, this implies a more advanced dissipation of supernova remnants, thus a lower local contribution to DM.
If ASKAP can only observe magnetars in an early stage, say $< 1000 ~\rm yr$, this could account for the high local contribution to DM, required above, but would also lower the inferred value of $\fIGM$.
Older magnetars potentially observed by CHIME would also cause a lower contribution to RM, which could be identified by statistical investigation of the distribution of RM observed by the different instruments.
However, the current sample of reported RM is too small to allow for any firm conclusions.
It might also be possible that separate frequency bands are dominated by completely different sources of FRBs.

Still, even with negligible local contributions, the CHIME results suggest a low IGM baryon content $\fIGM\lesssim 0.5$.
If this cannot be explained by false assumptions about the FRB redshift distribution and selection effects, the CHIME results would constitute strong evidence against missing baryons in the WHIM or, in fact, anywhere in the ionized IGM.


\subparagraph{Caveats}

Our conclusions are based on a number of assumptions regarding the IGM baryon content, the FRB redshift distribution as well as local contributions and foregrounds.
Furthermore, in our benchmark scenario we assumed all FRBs to be caused by magnetars in axisymmetric galaxies.
Alternative progenitor models beyond our assumptions might lead to different conclusions. 
For example, magnetars that only produce FRBs at an early stage, say less than $10^3 ~\rm yr$ after their formation, would produce much stronger dispersion than older ones ($\geq 10^5 ~\rm yr$), which we allowed in our benchmark scenario.
This would make lower values of $\fIGM$ more plausible, especially for the low-distance sample observed by ASKAP.
In future work, we will consider a wider range of assumptions in order to provide more significant results.
\\

Note that \frbpoppy~ uses estimates, e.g. of $\DM(z)$, in order to decide how many FRBs will be observed at a given redshift.
Theses estimates have been produced using slightly different assumptions on the contributing regions.
However, the $\DM$ is dominated by the IGM and the analytical description used in \frbpoppy~ provides a good match to our estimates.
Hence, we argue that this does not alter the general conclusions of this work. \\

Finally, in our benchmark scenario, we do not account for clustering or galaxy haloes \citep[e. g.][]{prochaska2019low,connor2020}.
These are especially important for the interpretation of FRBs located at low redshifts.
However, FRBs from higher redshift, for which these contributions are less important, are much more informative regarding $\pi(z)$ as well as $\fIGM$.
Our results are therefore reasonably robust against the additional presence of haloes along the line of sight.

%% file: conclusions.tex
\section{Conclusions}
\label{sec:conclusion}

Fast Radio Bursts are an important probe of Baryons in the Universe, as their dispersion measure is sensitive to all phases of the intergalactic medium, including the WHIM.
However, inference of the location of missing baryons with unlocalized FRBs is degenerate to the distribution of their host redshifts.
In this work we use the estimated extragalactic component of the dispersion measure to jointly infer  the redshift distribution of unlocalized FRBs and of the IGM baryon content with a Bayesian statistical approach. 
We  use an analytical model for the local environment of the source, assumed to be a magnetar, semi-analytic models for the ensemble of host and intervening galaxies, as well as models for the IGM from cosmological simulations. 
Comparing expectations to observations of unlocalized FRBs, we investigate the implications of events observed by ASKAP, CHIME and Parkes, listed in the FRBcat, using the open-source python software package \PreFRBLE\footnote{\href{https://github.com/FRBs/PreFRBLE}{github.com/FRBs/PreFRBLE}} \citep{PreFRBLEzenodo}.
Our main conclusions can be so summarised: 

\begin{itemize}
    \item From six localized FRBs beyond redshift $z\geq 0.1$, as listed in \citet{bhandari2020host}, we infer the intergalactic baryon content with 3$\sigma$ limits $\fIGM = 0.84^{+0.11}_{-0.39}$.
    FRBs thus potentially confirm the location of missing baryons in the WHIM. This value agrees well with results of \citet{li2020cosmologyinsensitive}.
    However, the error margin does not yet allow to confirm the WHIM solution beyond doubt.
    
    \item In order to unambiguously infer $\fIGM$ from unlocalized FRBs, the discrepancy between different instruments need to be resolved, e. g. by identifying different subsets of the FRB population. 
    A mere increase in the number of events will likely not solve this discrepancy.
    
    \item \new{In case that the WHIM solution to the missing baryon problem is correct, the Parkes sample is in conflict with a FRB population following the star formation.}

    \item Limited to the CHIME sample, we find that 
    the distribution which tracks the star formation rate is the least likely to explain observations.
    A better agreement is achieved for a distribution that peaks at rather low redshift $z \lesssim 0.6$, e. g. following the stellar mass density.
    
\end{itemize}

Though the small sample of localized FRBs suggest the missing baryons are in the ionized IGM, the error margin does not yet allow for unambiguous confirmation of this hypothesis.